# Design optimization of mode-matched bulk-mode piezoelectric micro-gyroscopes through modal analysis

Xiaosheng Wu[a,b], Wenyuan Chen[a], Abdolvand Reza[b]

[a]National Key Laboratory of Science and Technology on Nano/Micro Fabrication Technology, Research Institute of Micro/Nano Science and Technology, Shanghai Jiao Tong University, Dongchuan Road 800#, Shanghai 200240, China

[b]School of Electrical and Computer Engineering, Oklahoma State University, Tulsa, OK 74106, USA

**Abstract:** Bulk piezoelectric micro-gyroscope is a miniaturized inertial sensor that uses a differential thickness-shear bulk mode of a PZT block as the drive mode of the gyroscope. In the paper, a second differential thickness-extensional mode is identified for the sense mode and mode–matching is proposed for the first time by proper design of the device geometries. Through finite element modal analysis, the frequencies of drive mode and sense mode are obtained when the length of the PZT block varies from 4.8mm to 5.6mm and the width of the PZT block varies from 3.0mm to 4.0mm. Using a fitting method, the empirical formulae with an excellent fit are induced to predict the influence of the length and the width of the PZT block on the drive and sense mode frequencies. Based on these empirical formulae, the mode-matching equations are introduced. The analysis results show that for a given thickness of the PZT block, the effect of the width on the drive mode frequency is prominant. Conversely, the effect of length on the sense mode frequency is dominant. The resonance frequencies, kinetic energy ratios, scale factors of gyroscope are compared to evaluate the mode quality. The results show that the kinetic energy in y-axis direction of the drive mode and the kinetic energy in z-axis direction of the sense mode increase with the thickness of the PZT block, and consequently the scale factor of the gyroscope increases. For a constant thickness of the PZT block the scale factor will decrease as the length increases. Through design optimization we present a 20 times improvement in the scale factor of the mode-matched gyroscope. Given the thickness of PZT block, the length and the width will be determined by the mode-matching equations mentioned. Generally, the analysis suggests that the resolution of the gyroscope improves by increasing the thickness PZT block.

**Introduction**

Vibratory micro-gyroscopes are widely used in military, automotive, and consumer electronics. The examples include missile control, autopilot systems, active suspension sensors, crash sensors, image stabilization in digital cameras, and smart user interfaces in handheld devices.

Today, most of micro-machined vibrating gyroscopes are designed based on a discrete mass-spring structure, and they operate at relatively low frequencies ($\omega_0$=3-30kHz) [1,2]. If operated in 1-10mTorr vacuum, these micro-machined resonant gyroscopes can achieve quality factors values (Q) in the order of 50,000, which is mainly limited by thermoelastic damping (TED) in the flexing elements of the structure [3-5]. The fundamental mechanical Brownian noise of a Coriolis resonant gyroscope is given by [6]:

$$\Omega_{z(Brownian)} \propto \frac{1}{q_{drive}} \sqrt{\frac{4k_B T}{\omega_0 M Q_{Sense}}} \qquad (1)$$

Where $q_{drive}$ is the drive amplitude; $\omega_0$, $M$ and $Q_{Sense}$ are the natural frequency, mass and effective quality factor at the sense mode, respectively; $k_B$ is the Boltzmann constant and $T$ is the absolute temperature. The formula (1) is used for mode-matched micro-gyroscope and represents the effect of sense mode quality factor on improving the resolution. Based on this equation The drive resonant frequency of this micro-gyroscope is about 300KHz. According to (1), higher frequency and increased the efficient masses of the gyroscope resonant structure will work to improve the mechanical Brownian noise..

Maenaka et al have proposed the novel piezoelectric solid micro-gyroscope. The advantages of such a design compared to the traditional counterparts are its simple structure, robustness to crash or collision, and no requirement on vacuum package. It has been shown that in a higher order resonance mode, the vibrations for every point in the resonant structure are almost in one direction and differential on the adjacent edges of the prism and this special mode can be used as the primary mode [7-11]. The author proposed a piezoelectric solid micro-gyroscope with concentrated masses at corners of top and bottom surfaces of the resonant structure [8,9].

In the paper, we propose to further improve the performance of solid gyroscopes through mode-matching for the first time. At first, the working principle of this mode-matched piezoelectric micro-gyroscope is presented. Then, the modal analyses are described and based on the result of analysis, optimization of the mode-matched piezoelectric micro-gyroscope is concluded. The results show that the scale factor of the piezoelectric micro-gyroscope can be improved greatly through mode-matching.

**Working principle of piezoelectric bulk mode micro-gyroscope with mode-matched**

In resonant gyroscopes, two kinds of vibrations are utilized. One is the drive vibration (or the reference vibration), and the other is the sensing vibration. The energy of the sensing vibration is converted from the drive vibration due to the Coriolis effect induced by the rotation in the sensitive direction. Therefore, the frequency of sensing vibration is the same as that of the driving vibration. For mode-matched gyroscopes, the frequencies of the driving and sensing vibrations are matched with two resonant mode of the structure to increase the efficiency of energy conversion. Therefore, under mode-matched condition, the frequencies of drive mode and sense mode coincide. For mode-matched vibratory gyroscopes, the sensing vibration is magnified by its quality factor to improve its resolution [12].

However, in practice the frequencies of the two modes are not going to end up perfectly matching either as a result of inevitable process errors and/or relative shift of frequency as a function of environmental parameters such as temperature. Therefore, the Q amplification effect can quickly vanish or the output of the gyroscope becomes very sensitive to environmental parameters if the quality factor of the sense mode is high. This can be better understood by considering equation (2) through which it is shown that the half-power bandwidth $\Delta f$ is proportional to the resonance frequency $f_0$ of sensing vibration and is inversely proportional to quality factor $Q$:

$$\Delta f = \frac{f_0}{Q} \qquad (2)$$

In order to mitigate this limitation several approaches have been proposed including gyroscopes with multiple sense modes that result in a filter-like frequency responses with increased range of frequency for which the Q-amplification is relatively constant [13]. Also, there are some micro-machined vibratory gyroscopes whose sensing vibration works near the resonance mode or not in the resonance mode to increase its working bandwidth [14,15]. For the bulk piezoelectric micro-gyroscope, the mechanical damping is relatively large for PZT material, and its material quality factor $Q$ is normally not larger than 100. Also, the operation frequency of bulk piezoelectric micro-gyroscope is larger than that of conventional micromechanical vibratory gyroscopes. Considering these two factors, the bulk piezoelectric micro-gyroscope can operate in the mode-matched condition without significant loss of bandwidth.

Fig.1 schematically shows the working principle of bulk piezoelectric micro-gyroscope. The PZT rectangular block is polarized along the z-axis. The drive mode of the bulk piezoelectric micro-gyroscope is a differential thickness shear resonance mode, in which the strain will change polarity along the thickness and the length of the block The differential drive vibration along the y-axis (u) is shown in fig1. When the driving voltage are applied on electrodes D+ and D-, the drive vibration is excited in the PZT resonator. If there is no rotation input, the voltage amplitudes on the sensing electrodes S1, S2, S3, S4 are equal. The reference electrodes R1, R2 are used to monitor the drive vibration. When there is rotation input in the y-axis direction, the Coriolis forces are induced in the z-axis direction. Then, voltage amplitudes on the sensing electrodes are no longer equal and their differential values will develop that correlate to the rotation input in the y-axis direction. The electrodes on the bottom surface of PZT resonator are the same as those on the top surface.

Through finite element analysis of PZT rectangular parallelepiped, The mode shape for the drive and the sense mode are identified and shown in Fig. 2(a) and 2(b) respectively. For the drive mode, the strain is mainly in the y-axis direction, whereas for the sense mode strain is mainly in the z-axis. This is consistent with the working principle of the bulk piezoelectric micro-gyroscope shown in Fig.1. In order to get the mode-matched, the resonant frequency of both vibration modes should equal. The factors which affect resonant frequencies of drive mode and sense mode include dimension of the PZT block and the material properties. The PZT material presumed in this paper is PZT-5H. We will study the influence of resonator sizes on the mode matching for this kind of gyroscope. This includes the effect of length (L), width (W) and thickness (T) of the PZT resonator on its drive mode frequency and sense mode frequency. Then, we can find the requirement of mode-matched on the sizes of the resonator.

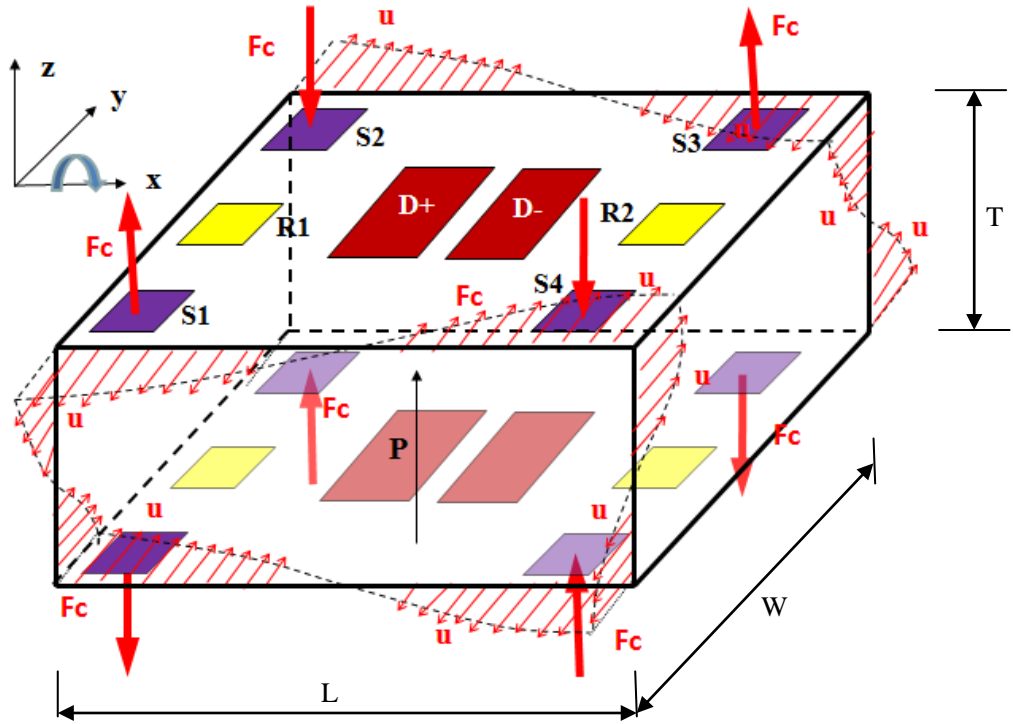

Fig.1 working principle of bulk piezoelectric micro-gyroscope (D+, D-: driving electrodes; S1, S2, S3, S4: sensing electrodes; R1, R2: reference electrodes; P: poling direction of PZT; Fc: Coriolis forces; u: resonant displacement)

For the perfect drive vibration or sensing vibration, the points in the PZT block only vibration in only one direction. As shown in fig. 2, for the drive vibration, the vibration direction is in the y-axis direction, and for the sensing vibration, the vibration direction is in the z-axis direction. However, the spurious modes, which make the PZT block vibrate in other directions, always exist. From fig.2(a), it is not difficult to notice the spurious mode of thickness flexure. Through the optimized design in the paper, the spurious mode is minimized to improve performance of the gyroscope.

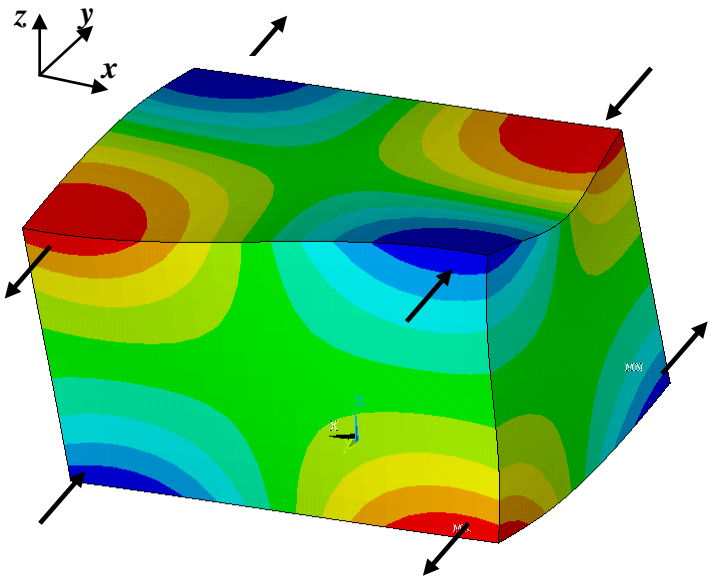 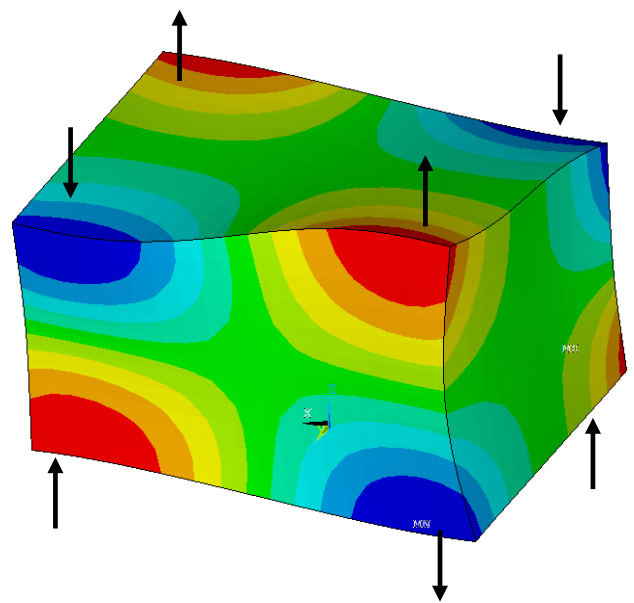

Fig 2(a) drive mode          Fig.2(b) sense mode

Fig.2 The color of contour represents the displacement in y axis direction

The material properties of PZT-5H (poling in z direction) used are as follows: elastic coefficients $s_{11}=16.5\times10^{-12}$ $m^2/N$, $s_{12}=-4.78\times10^{-12}$ $m^2/N$, $s_{13}=-8.45\times10^{-12}$ $m^2/N$, $s_{33}=20.7\times10^{-12}$ $m^2/N$, $s_{55}=43.5\times10^{-12}$ $m^2/N$, $s_{66}=42.6\times10^{-12}$ $m^2/N$;

piezoelectric strain constant $d_{31}$=-274×10$^{-12}$ m/V, $d_{33}$=593×10$^{-12}$ m/V, $d_{15}$=741×10$^{-12}$ m/V; relative electric permittivity $\varepsilon_{11}$=3130, $\varepsilon_{33}$=3400.

According to the piezoelectric theory, the constitutive relation of electric variants and mechanic variants is set through these material parameters. The constitutive relations is shown in formula (3).

$$S = s^E \cdot T + d^t \cdot E$$
$$D = d \cdot T + \varepsilon^T \cdot E \qquad (3)$$

S: strain vector;

T: stress vector;

D: electric charge density displacement vector;

E: electric field vector;

s: piezoelectric compliance matrix;

ε: electric permittivity matrix;

d: piezoelectric coupling coefficients matrix for strain-charge form;

**Modal analysis of bulk piezoelectric micro-gyroscope**

In the modal analysis, we select ANSYS as the finite element tool. Piezoelectrics is the coupling of structural and electric fields, which is a natural property of some crystals (quartz, etc) and ceramics (PZT, etc). In ANSYS, the elements of SOLID5, SOLID98, SOLID226 and SOLID227 can be used as the 3D piezoelectric multiphysics analysis. Here, we select SOLID227 as the element type. Comparing with other elements, SOLID226 is a kind of higher order element, which has 20 nodes for each element. Higher precision can be obtained through using SOLID226 element. SOLID226 has strong coupling function. Besides piezoelectric coupling, it can also include Coriolis effect in the analysis, and this provides convenience for the further gyroscope resolution analysis.

Concerning the small size of PZT resonator, we will change the length from 4.8mm to 5.6mm, and the width from 3.0mm to 4.0mm, and the thickness from 2.8mm to 3.2mm.

In the modal analysis, the meshing result of the PZT rectangular parallelepiped is shown in Fig.3. We decreased the element size small enough to get the convergence results. Here, the length and width of element are selected as 0.25mm, and the height is 0.1mm. The elements and the nodes number depend on the size of the PZT resonator.

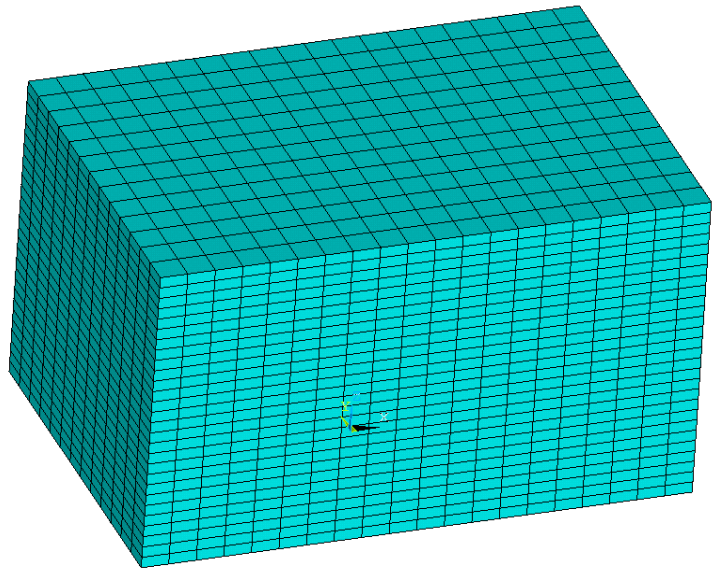

Fig.3 meshing result of finite element model (with length: 4.8mm, width: 3.21, height: 2.8mm, elements number: 7280, nodes number: 32795)

Fig.4 gives the results of dependence of drive mode (shown as fig.2(a)) frequency on length and width of PZT block. From Fig.4(a), we can see that as the length of PZT block increases, the drive mode frequency decreases. However, the frequency decrease rate is very slow. From Fig.4(b), we can see that as the width of PZT block increases, the drive mode frequency decreases relatively quickly. For this analysis, the thickness of the PZT block is set to be 3mm.

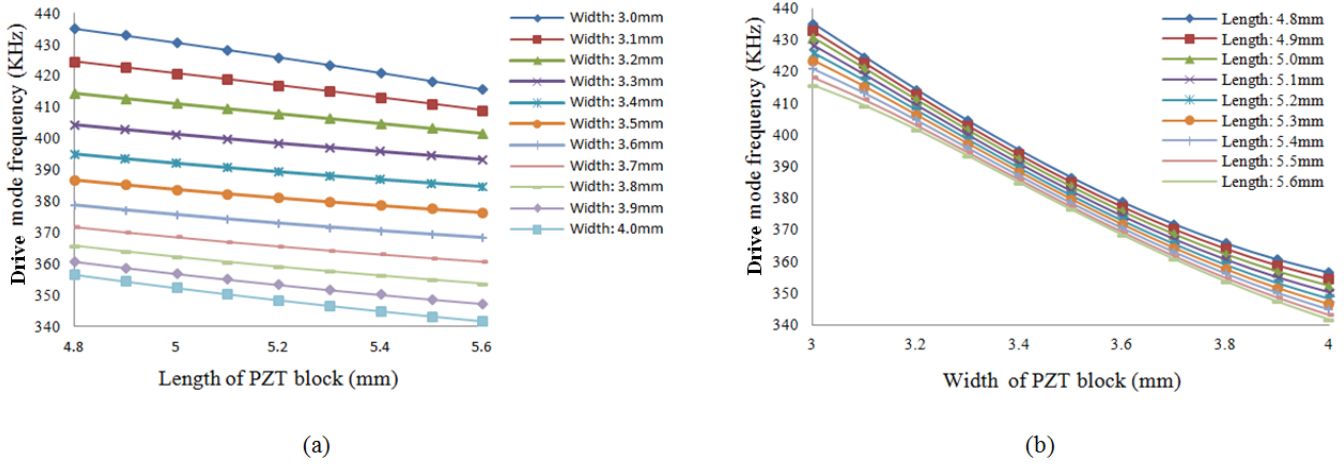

Fig.4 Dependence of drive mode frequency on length and width of PZT block

From fig.4, we can find that for a given thickness of PZT block, the drive mode frequency is only dependent on length and width of the block. In order to obtain a general relation between the drive mode frequency ($f_{ref}$) and length ($l$), width ($w$) of the PZT block, we use the cubic polynomial to fit this dependent relation, as shown in formula (4).

$$f_{\text{ref}} = 1805 - 458.7w - 294.2l - 50.83w^2 + 176.7wl - 8.746l^2 + 24.47w^3 - 35.03w^2l + 6.938wl^2 - 0.8542l^3 \quad (4)$$

Here, $w$ and $l$ refer to width and length of PZT block respectively.

The goodness of fit is[16]:

SSE (Sum of Squares due to Error): 4.591; R-square: 0.9999; RMSE (Root Mean Squared Error): 0.2359

Formula (4) gives a very accurate estimate of drive mode frequency when thickness of PZT block is 3.0mm.

Fig.5 gives the dependence of sense mode (shown as fig.2(b)) frequency on length and width of PZT block. From Fig.5(a), we can see that as the length of PZT block increases, the sense mode frequency decreases relatively quickly. From Fig.5(b), we can see that as the width of PZT block increases, the sense mode frequency increases, but the frequency increasing rate is slow. For this analysis, the height of the PZT block is set to be 3mm.

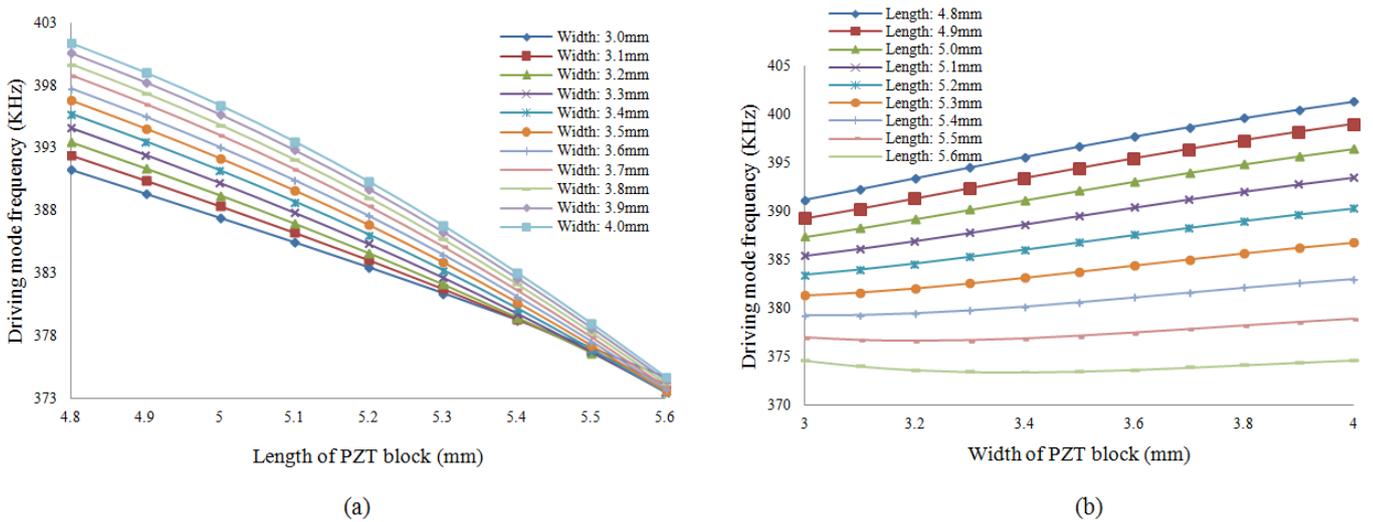

Fig.5 Dependence of sense mode frequency on length and width of PZT block

Same as that of drive frequency, we also get the fitting relation between length ($l$), width ($w$) and sense mode frequency using cubic polynomial, as shown in formula (5).

$$f_{\text{sense}} = 946.2 - 69.97w - 321.4l - 5.602w^2 + 48.79wl + 53.91l^2 - 3.23w^3 + 7.753w^2l - 11.11wl^2 - 1.595l^3 \quad (5)$$

The goodness of fit is:

SSE (Sum of Squares due to Error): 0.2496;   R-square:1;   RMSE (Root Mean Squared Error): 0.05296

In order to meet the mode matching condition, then

$$f_{\text{ref}} = f_{\text{sense}} \tag{6}$$

Taking formula (4) and (5) into (6), equation (7) is obtained.

$$858.8-388.73w+27.2l-45.228w^2+127.91wl-62.656l^2+27.7w^3-42.7830w^2l+18.048wl^2+0.7408l^3=0 \tag{7}$$

Formula (7) gives the condition that should be met to get mode-matched gyroscope when thickness of the piezoelectric block is 3mm. Because formula (7) is derived from the simulation data when length is ranged from 4.8 to 5.6mm, and width is ranged from 3.0 to 4.0mm, the estimate error will become large when length or width of the PZT block is out of their range mentioned.

Using the same method, we obtained the mode-matched conditions of bulk piezoelectric micro-gyroscope when thickness of the piezoelectric block is 2.8mm (formula (8)), 2.9mm(formula(9)), 3.1mm(formula(10)), 3.2mm(formula(11)) respectively.

$$3641-854.1w-1321.2l-38.795w^2+297.25wl+145.3l^2+27.913w^3-42.45w^2l+0.554wl^2-8.5901l^3=0 \tag{8}$$
$$1883.6-592.33w-439.4l-31.687w^2+186.45wl+7.47l^2+26.955w^3-42.97w^2l+12.211wl^2-2.2741l^3=0 \tag{9}$$
$$370-303.8w+247.7l-48.57w^2+102.92wl-94.51l^2+27.173w^3-41.868w^2l+19.999wl^2+2.0925l^3=0 \tag{10}$$
$$406+361.2w-722.87l+19.20w^2-89.88wl+178.79l^2-25.02w^3+43.775w^2l-22.48wl^2-6.714l^3=0 \tag{11}$$

The equation of formula(7), (8), (9), (10 ) and (11) can be plotted in one figure, as shown in Fig.6. In figure, any point on each curve represents a couple of width and length value, for which the matched-mode can be obtained. Each curve corresponds to a kind of PZT block thickness. From the figure, we can see that as the thickness of PZT block is larger, the curve become steeper. That is to say, as the thickness is larger, the mode-matched condition become more sensitive with width value. In order to evaluate the mode quality, we select 3 points from each curve, and research the mode quality for each point.

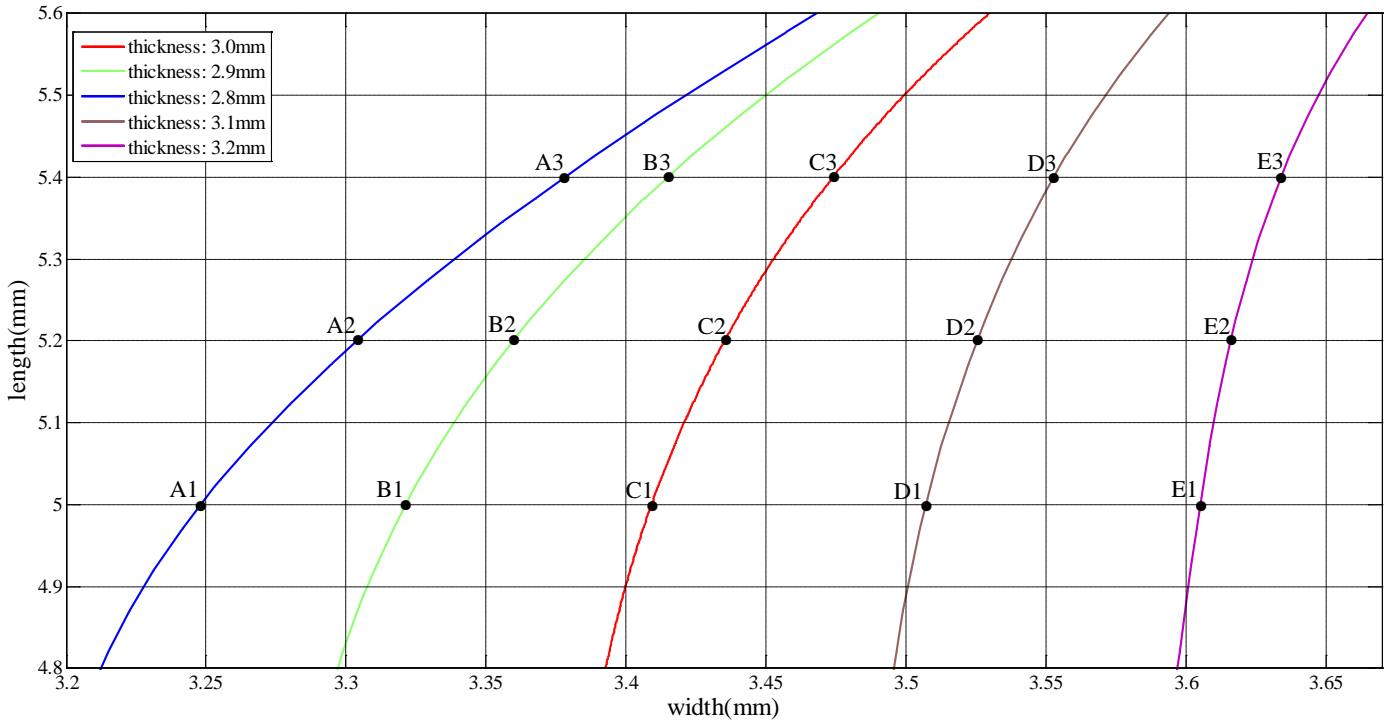

Fig.6 mode-matched condition of width and length of PZT block

**Evaluation of modes quality of drive vibration and sensing vibration.**

As shown in fig.6, the PZT block with the size corresponding to each point in the curve can get the drive and sense mode-matched. From the analysis of the gyroscope principle, we know that for the perfect vibration modes, the vibration in drive mode and sense mode are in only one direction. However, there are always spurious modes coupled in the drive mode and sense mode. The spurious modes usually vibrate in other directions different from the direction of drive mode or sense mode. These spurious modes influence the mode quality of drive vibration and sensing vibration. From the aspect of kinetic energy, we analyzed the modes quality of drive vibration and sensing vibration.

We defined three indicators to evaluate the kinetic energy ratio in three orthogonal directions, as shown in formulae (12)-(14). Here, $K_{Ex}$, $K_{Ey}$, $K_{Ez}$, are the kinetic energy in the x-axis, y-axis and z-axis direction respectively. $K_{Et}$ is the total kinetic energy. $R_{Ex}$, $R_{Ey}$, $R_{Ez}$ are kinetic energy ratio in three orthogonal directions. For the drive mode with higher quality, $R_{Ey}$ should be larger. For the sense mode with higher quality, $R_{Ez}$ should be larger.

$$R_{Ex} = \frac{K_{Ex}}{K_{Et}} \tag{12}$$

$$R_{Ey} = \frac{K_{Ey}}{K_{Et}} \tag{13}$$

$$R_{Ey} = \frac{K_{Ez}}{K_{Et}} \tag{14}$$

$$K_{Et} = K_{Ex} + K_{Ey} + K_{Ez} \tag{15}$$

$$K_{Ex} = \sum_{i=1}^{N} \frac{1}{2} m_i v_{xi}^2 \tag{16}$$

N: number of element in the finite element analysis, $m_i$: mass of the $i$th element, $v_{xi}$: velocity in x-axis direction of the $i$th element.

$$u_{xi} = U_{xi} \cdot \sin(\omega t) \tag{17}$$

$u_{xi}$: displacement in x-axis direction of the $i$th element, $U_{xi}$: vibration amplitude in x-axis direction of the $i$th element, $\omega$: vibration angular frequency

$$v_{xi} = \frac{du_{xi}}{dt} = U_{xi} \cdot \omega \cdot \cos(\omega t) \tag{18}$$

$$K_{Ex} = \sum_{i=1}^{N} \frac{1}{2} m_i (U_{xi} \cdot \omega \cdot \cos(\omega t))^2 \tag{19}$$

$$K_{Ey} = \sum_{i=1}^{N} \frac{1}{2} m_i (U_{yi} \cdot \omega \cdot \cos(\omega t))^2 \tag{20}$$

$$K_{Ez} = \sum_{i=1}^{N} \frac{1}{2} m_i (U_{zi} \cdot \omega \cdot \cos(\omega t))^2 \tag{21}$$

As shown in fig.3, every element has equal mass due to map meshing. Then, from (12), (13), (14) and (15), we get

$$R_{\mathrm{Ex}} = \frac{\sum_{i=1}^{N}(U_{xi})^2}{\sum_{i=1}^{N}(U_{xi})^2 + \sum_{i=1}^{N}(U_{yi})^2 + \sum_{i=1}^{N}(U_{zi})^2} \tag{22}$$

$$R_{\mathrm{Ey}} = \frac{\sum_{i=1}^{N}(U_{yi})^2}{\sum_{i=1}^{N}(U_{xi})^2 + \sum_{i=1}^{N}(U_{yi})^2 + \sum_{i=1}^{N}(U_{zi})^2} \tag{23}$$

$$R_{\mathrm{Ez}} = \frac{\sum_{i=1}^{N}(U_{zi})^2}{\sum_{i=1}^{N}(U_{xi})^2 + \sum_{i=1}^{N}(U_{yi})^2 + \sum_{i=1}^{N}(U_{zi})^2} \tag{24}$$

$U_{xi}$, $U_{yi}$, $U_{zi}$ can be extracted from finite element results.

As shown in fig.6, we select 3 points from each curve to evaluate their mode quality. The PZT block size and resonant frequency corresponding to each point is listed in table 1.

Table 1 size, resonant frequency, kinetic energy ratio, gyroscope scale factor of typical points on fitting mode matched curve

| Point | Size(mm) | $f_r, f_s$ (KHz)[*] | Kinetic energy ratio[**] | SF(μV/rad/s)[***] |
|---|---|---|---|---|
| A1 | L:5.0, W:3.248, T:2.8 | 408.07, 408.10 | $R_{\mathrm{Ey}}$:0.4202, $R_{\mathrm{Ez}}$:0.3412 | 82.23 |
| A2 | L:5.2, W:3.303, T:2.8 | 400.27, 400.23 | $R_{\mathrm{Ey}}$:0.4060, $R_{\mathrm{Ez}}$:0.2703 | 70.66 |
| A3 | L:5.4, W:3.378, T:2.8 | 391.07, 391.08 | $R_{\mathrm{Ey}}$:0.3949, $R_{\mathrm{Ez}}$:0.2035 | 63.78 |
| B1 | L:5.0, W:3.321, T:2.9 | 400.20, 400.20 | $R_{\mathrm{Ey}}$:0.4330, $R_{\mathrm{Ez}}$:0.3938 | 81.80 |
| B2 | L:5.2, W:3.360, T:2.9 | 394.97, 394.01 | $R_{\mathrm{Ey}}$:0.4191, $R_{\mathrm{Ez}}$:0.3342 | 77.71 |
| B3 | L:5.4, W:3.415, T:2.9 | 386.70, 386.72 | $R_{\mathrm{Ey}}$:0.4066, $R_{\mathrm{Ez}}$:0.2658 | 69.86 |
| C1 | L:5.0, W:3.409, T:3.0 | 391.26, 391.26 | $R_{\mathrm{Ey}}$:0.4443, $R_{\mathrm{Ez}}$:0.4317 | 88.25 |
| C2 | L:5.2, W:3.435, T:3.0 | 386.28, 386.27 | $R_{\mathrm{Ey}}$:0.4308, $R_{\mathrm{Ez}}$:0.3859 | 86.70 |
| C3 | L:5.4, W:3.474, T:3.0 | 380.47, 380.48 | $R_{\mathrm{Ey}}$:0.4177, $R_{\mathrm{Ez}}$:0.3269 | 80.32 |
| D1 | L:5.0, W:3.507, T:3.1 | 382.41, 382.41 | $R_{\mathrm{Ey}}$:0.4551, $R_{\mathrm{Ez}}$:0.4600 | 85.23 |
| D2 | L:5.2, W:3.525, T:3.1 | 378.24, 378.23 | $R_{\mathrm{Ey}}$:0.4428, $R_{\mathrm{Ez}}$:0.4256 | 88.44 |
| D3 | L:5.4, W:3.553, T:3.1 | 373.47, 373.50 | $R_{\mathrm{Ey}}$:0.4289, $R_{\mathrm{Ez}}$:0.3794 | 87.75 |
| E1 | L:5.0, W:3.605, T:3.2 | 372.91, 372.91 | $R_{\mathrm{Ey}}$:0.4653, $R_{\mathrm{Ez}}$:0.4804 | 90.14 |
| E2 | L:5.2, W:3.616, T:3.2 | 369.25, 269.26 | $R_{\mathrm{Ey}}$:0.4546, $R_{\mathrm{Ez}}$:0.4539 | 89.20 |
| E3 | L:5.4, W:3.633, T:3.2 | 365.35, 365.27 | $R_{\mathrm{Ey}}$:0.4418, $R_{\mathrm{Ez}}$:0.4190 | 90.56 |

[*]: $f_r$, drive resonant frequency calculated from fit formulae; $f_s$, sensing resonant frequency calculated from fit formulae

[**]: $R_{\mathrm{Ey}}$, ratio of kinetic energy caused by y-axis velocity to total kinetic energy in the drive mode vibration; $R_{\mathrm{Ez}}$, ratio of kinetic energy caused by z-axis velocity to total kinetic energy in the sense mode vibration

[***]: scale factor calculating from the voltage difference on the two adjacent sensing electrodes, the sizes of driving electrodes are same, and the driving voltage amplitude is same.

Table 1 gives the characteristic parameters of typical points on the mode-matched curves shown in fig.6, and these parameters include size of PZT block, the drive resonant frequency and the sensing resonant frequency, y-axis kinetic energy ratio of drive mode vibration and z-axis kinetic energy ratio of sense mode, scale factor of gyroscope. Analyzing the data shown, we can find that when length is same, as the thickness of PZT block becomes larger, $R_{\mathrm{Ey}}$, $R_{\mathrm{Ez}}$ and scale factor become larger. Here, scale factor refers to the voltage difference on two adjacent sensing electrodes when 1 rad/s rotation is input in the sensitive direction. These validate the analysis in the gyroscope principle. That is to say that increasing the portion of drive vibration in y-axis direction and the portion of sensing vibration in z-axis direction will improve the gyroscope precision. For all of these points, the drive modal frequency and sensing modal frequency almost equal, so the drive and sense mode-matched is obtained for these points. When the thickness of PZT block is same, increasing the length of PZT block will decrease the portion of drive vibration in y-axis direction and the portion of sensing vibration in z-axis direction, and this will decrease the scale factor the gyroscope, especial when thickness is smaller.

**Conclusion**

Bulk piezoelectric micro-gyroscope is a miniaturized inertial sensor that uses a thickness-shear mode of PZT block as the drive mode of the gyroscope. In the paper, a second thickness-extension mode is found as the sense mode, and mode-matched concept for this bulk piezoelectric micro-gyroscope is firstly proposed. Through finite element modal analysis, the frequencies of drive mode and sense mode are obtained when length of PZT block changes from 4.8mm to 5.6mm and the width of PZT block changes from 3.0mm to 4.0mm. Using fitting method, the empirical formulae with an excellent fit are induced to predict the influence of length, width of PZT block on the drive mode frequency and sense mode frequency. Based on the empirical formulae of drive mode frequency and sense mode frequency, the mode-matched equations are introduced. Five mode-matched equations are obtained in the paper, each standing for thickness of PZT block, 2.8mm, 2.9mm, 3.0mm, 3.1mm, 3.2mm respectively. The effect of the width on the drive mode frequency is prominent. Conversely, the effect of length on the sense mode frequency is dominant. Therefore, given the thickness of PZT block, the width influences drive frequency more, and the length influences sensing frequency more.

In order to evaluate the mode-matched quality, 15 points are picked up from the five mode matched relation curves, and their resonance frequencies, kinetic energy ratios, scale factors of gyroscope are compared. The results indicate that larger thickness of the PZT block is good for increasing the kinetic energy in y-axis direction of drive mode and the kinetic energy in z-axis direction of sense mode, and then increase the scale factor of the gyroscope further. Given the thickness of PZT block, larger length will decrease the scale factor.

We also calculated the PZT block with the size (5mm length×4mm width×3mm thickness) without mode-matched. With this size, the drive modal frequency is 352.252KHz and the sensing modal frequency is 396.434KHz. With the same sizes of driving electrodes and the same driving voltage amplitude, we got the scale factor 4.59μV/rad/s at drive modal frequency of 352.252KHz. Comparing with the simulated results of the mode-matched model shown in table 1, the scale factor of the mode-matched gyroscope has improved nearly 20 times better.

Structure of piezoelectric bulk mode micro-gyroscope is very simple, and it is just a PZT block. However, for the mode-matched micro-gyroscope of this kind, it has very strict requirement on its size. Given the thickness of PZT block, the length and width will be determined according to the mode-matched condition. The mode-matched condition also indicates that the thicker of the PZT block is, the condition is more sensitive with the width. This will make the fabrication more difficult.

The performance of the piezoelectric bulk mode micro-gyroscope is also influence by the shape and size of electrodes distributed on the top and bottom surfaces of PZT block. This will be considered in the further research.

**Acknowledgments** This work was supported by Education Department of China Supporting Project (625010117) and China Scholarship Council (No.2010831025).

**Reference**

[1] Yazdi N, Ayazi F, Najafi K, (1998) Micromachined inertial sensors. Proceedings of the IEEE, 86: (8) pp1640-1659

[2] Liu K, Zhang WP, Chen WY, Li K, Dai FY, Cui F, Wu XS, Ma GY, Xiao QJ, (2009) The development of micro-gyroscope technology. J. Micromech. Microeng., 29: pp1-29

[3] Duwel A, Candler RN, Kenny TW, Varghese M, (2006) Engineering MEMS resonators with low thermaoelastic Damping. Journal of Microelectromechanical Systems, 15: (6) pp1437-1445

[4] Hao ZL, Ayazi F, (2005) Thermoelastic damping in flexural mode ring gyroscopes. 2005 ASME, Orlando, FL, Nov. 5-11

[5] Kim B, Hopcroft MA, Candler RN, Jha CM, Agarwal M, (2008) Temperature dependence of quality factor in MEMS resonators. Journal of Microelectromechanical Systems, 17: (3) pp755-766

[6] Sharma A, Zaman M, Ayazi F, (2007) A 104-dB dynamic range transimpedance-based CMOS ASIC for tuning fork microgyroscopes. IEEE J. Solid-State Circuits, 42: (8) pp1790-1802

[7] Maenaka K, Kohara H, Nishimura M, Fujita T and Takayama Y (2006) Novel solid micro-gyroscope. Tech. Dig. of the 19th IEEE Int. Conf. on Microelectromechanical Systems (Istanbul) pp 634–7

[8] Wu XS, Chen WY, Lu YP, Zhang WP, (2009) Modeling analysis of piezoelectric micromachined modal gyroscope (PMMG). Proceedings of the 2009 4th IEEE International Conference on Nano/Micro Engineered and Molecular Systems (Shenzhen) pp304-309

[9] Wu XS, Chen WY, Lu YP, Xiao QJ, Ma GY, Zhang WP, Cui F (2009) Vibration analysis of a piezoelectric micromachined modal gyroscope (PMMG). J. of Micromech Microeng 19: (12) 125008

[10] Lu YP, Wu XS, Zhang WP, (2010) Optimization and analysis of novel piezoelectric solid micro-gyroscope with high resistance to shock. Microsystem Technologies, 16: (4) pp571-584

[11] He YS, Wu XS, Zheng F, Chen WY, Zhang WP, Cui F, Liu W, (2013) Closed loop driving and detect circuit of piezoelectric solid-state micro-gyroscope. Microsystem Technologies, Published online

[12] Zaman MF, Sharma A, Ayazi F, (2006) High performance matched-mode tuning fork gyroscope. Tech. Dig. of the 19th IEEE Int. Conf. on Microelectromechanical Systems (Istanbul) pp66-69

[13] Acar C, Shkel AM, (2005) An approach for increasing drive-mode bandwidth of MEMS vibratory gyroscopes. Journal of Microelectromechanical Systems, 14: (3) pp520-528

[14] Acar C, Schofield AR, Trusov AA, Costlow LE, Shkel AM, (2009) Environmentally robust MEMS vibratory gyroscopes for automotive applications. Journal of IEEE Sensors, 9: 1895-1906

[15] Si C, Han G, Ning J, Yang F, (2013) Bandwidth optimization design of a multi degree of freedom MEMS gyroscope. Sensors, 13: 10550-10560

[16] http://w3eos.whoi.edu/12.747/notes/lect03/lectno03.html